\documentclass[twocolumn]{aastex63}

\accepted{on June 11, 2020, to publish in The Astronomical Journal}

\shorttitle{Feasibility of Observing Gamma-ray Polarization from Cygnus X-1 Using a CubeSat}
\shortauthors{Yang et al.}

\begin{document}

\title{Feasibility of Observing Gamma-ray Polarization from Cygnus X-1 Using a CubeSat}

\correspondingauthor{Hsiang-Kuang Chang}
\email{hkchang@mx.nthu.edu.tw}

\author{Chien-Ying Yang}
\affiliation{Institute of Astronomy, National Tsing Hua University, Hsinchu, Taiwan}
\author{Yi-Chi Chang}
\affiliation{Institute of Astronomy, National Tsing Hua University, Hsinchu, Taiwan}
\author{Hung-Hsiang Liang}
\affiliation{Institute of Astronomy, National Tsing Hua University, Hsinchu, Taiwan}
\author{Che-Yen Chu}
\affiliation{Institute of Astronomy, National Tsing Hua University, Hsinchu, Taiwan}
\author{Jr-Yue Hsiang}
\affiliation{Institute of Astronomy, National Tsing Hua University, Hsinchu, Taiwan}
\author{Jeng-Lun Chiu}
\affiliation{National Space Organization, National Applied Research Labs, Hsinchu, Taiwan}
\author[0000-0002-8578-7775]{Chih-Hsun Lin}
\affiliation{Institute of Physics, Academia Sinica, Taipei, Taiwan}
\author{Philippe Laurent}
\affiliation{CEA/DRF/IRFU/DAp, Saclay, France}
\author{Jerome Rodriguez}
\affiliation{CEA/DRF/IRFU/DAp, Saclay, France}
\author[0000-0002-5617-3117]{Hsiang-Kuang Chang}
\affiliation{Institute of Astronomy, National Tsing Hua University, Hsinchu, Taiwan}
\affiliation{Department of Physics, National Tsing Hua University, Hsinchu, Taiwan}

\begin{abstract}
Instruments flown on CubeSats are small. Meaningful applications of CubeSats in astronomical observations rely on the choice of a particular subject that is feasible for CubeSats.
Here we report the result of a feasibility study for observing gamma-ray polarization from Cygnus X-1 using a small Compton polarimeter on board a 3U CubeSat.
Silicon detectors and cerium bromide scintillators were employed in the instrument models that we discussed in this study. 
Through Monte Carlo simulations with Geant4-based MEGAlib package, we found that, 
with a 10-Ms on-axis, zenith-direction
 observation in a low inclination, 
low altitude earth orbit radiation background environment, 
the minimum detectable polarization degree can be down to about 10\% in 160 - 250 keV, 20\% in 250 - 400 keV, and 65\% in 400 - 2000 keV.
A 3U CubeSat dedicated to observing Cygnus X-1 can therefore yield useful information on the polarization state of  gamma-ray emissions from the brightest persistent X-ray black-hole binary in the sky.
\end{abstract}

                                                     
\section{Introduction} \label{sec:intro}
Because of much lower cost and much shorter developing cycle than traditional space missions with larger satellites,
CubeSat missions have been blooming in recent years. A unified specification for the bus is forming. 
According to CubeSat Design Specification (CDS, \url{http://www.cubesat.org} ),
one unit of CubeSat (1U) is 10 cm $\times$ 10 cm $\times$ 11.35 cm in size and weighs about 0.8 -- 1.3 kg. 
A CubeSat can be a combination of several units. CubeSats have been utilized in many different fields, including
astronomy and astrophysics \citep{shkolnik18}. 
Among others, a small gas pixel detector, housed in 1U of a 6U CubeSat, to measure X-ray polarization of the Crab at keV regime is currently in operation (PolarLight, \citet{feng19,feng20}).
An MeV telescope on a CubeSat has also been discussed in the literature \citep{lucchetta17,rando19}.

Cygnus X-1 (Cyg X-1) is the brightest persistent X-ray black-hole binary (BHB) in the sky.  
Its black hole mass is estimated to be about 15 M$_\odot$ and the companion is a supergiant star of more than 20 M$_\odot$ \citep{ziolkowski14}.
In X-ray and soft gamma-ray bands, its emission mainly consists of a thermal component below about 10 keV, a Comptonization component between
10 keV and several hundred keV, and another power-law component (sometimes undetected during the soft state) at even higher energy.
The low energy thermal one is believed to come from the accretion disk and the Comptonization one is due to the reprocessing of photons from the disk
by higher-energy particles  in certain Compton clouds. Reflection of these reprocessed (Comptonized) photons from the Compton clouds by the accretion disk is often needed to better understand the spectrum in this energy range. Based on measurements of IBIS and SPI on board INTEGRAL,
the high-energy power-law component was reported to show
high degree of linear polarization \citep{laurent11,jourdain12,rodriguez15}, 
which implies a magnetic field strength stronger than the equipartition value in the jet, if the emission is due to synchrotron radiation from the jet
\citep{zdziarski14}. The emission between 250 -- 400 keV, which contains Comptonized photons (plus reflection) and some contribution from the jet,
has a somewhat smaller polarization degree of $(40\pm 10$)\%, based on SPI measurement \citep{jourdain12}, but undetected by IBIS with a 20\% upper limit.  At even lower energies, SPI reported a 20\% upper limit for 130 -- 230 keV \citep{jourdain12} and PoGO+ gave an 8.6\% upper limit for 20 -- 180 keV \citep{chauvin18}, which argues for an extended, rather than compact, Compton cloud region \citep{chauvin18, chauvin19}. 
Polarization of $(2.4\pm 1.1)$\% at 2.6 keV and $(5.3\pm 2.5)$\% at 5.2 keV was also reported by OSO-8 \citep{long80}.

In view of scientific importance of measuring polarization states of soft gamma-ray emissions from Cyg X-1, we propose to build
a small Compton polarimeter to fly on a 3U CubeSat.
In this paper we report the result of the feasibility of such a concept, based on simulations using MEGAlib \citep{zoglauer08}.
We describe instrument models in Section \ref{sec:model},  their performance in Section \ref{sec:perform}, and the expected
minimum detectable polarization (MDP) in Section \ref{sec:mdp}.

\section{Instrument models} \label{sec:model}

In a CubeSat, because of its small size, a COMPTEL-like time-of-flight configuration is not practical.
A compact Compton telescope fits better.
One example of a compact Compton telescope is the Compton Spectrometer and Imager (COSI) \citep{kierans17,yang18}, in which
a cross-strip high-purity germanium detector array is employed. 
High-purity germanium detectors require cooling systems to make the working temperature down to about 80 K.
It is therefore not feasible either for a CubeSat.
Instead, silicon sensors and scintillators are more suitable, 
such as that proposed by \citet{rando19} and employed in BurstCube \citep{smith19}.
In the attempt of building a Compton polarimeter to fly on a CubeSat, we first considered a combination of silicon sensors and cerium bromide (CeBr$_3$) scintillators, with the former
acting mainly as the scatter and the latter mainly as the absorber.
 
From first principles of Compton kinetics and cross-sections, 
we first designed a sensitive telescope specially optimized for polarization measurement, 
which is not the case for the current space missions, such as ASTROSAT or INTEGRAL. 
In general, such a Compton polarimeter is made of stacked detectors, 
where the photon is first scattered in the first layer and then absorbed in the detector below 
(cylindrical configuration where the scattering detectors is surrounded by
the absorbing ones, as in the IKAROS/GAP experiment  \citep{yonetoku11} is also feasible, 
but is more complicated to implement in a CubeSat). 
We can then derive the source'€™s polarization properties (polarization angle and fraction) 
through the study of the azimuthal distribution of the detected photons on the second detector. 
To do so, we should measure the position X,Y and energy of each photon detected in the stacked detectors. 
These detectors should be thus spectro-imagers.

The first layer of the telescope should favor Compton scattering, thus should be derived from a low Z material. 
On the other hand, the second detector should absorb the scattered photons and thus a high Z material is preferable. 
A low Z spectro-imager could be realized using a pixelated or stripped silicon detector. 
The high Z detector could be either a semi-conductor (CdTe for instance) 
or a scintillator with a sufficient light yield to ensure a good energy and position measurement. 
LaBr$_3$ and CeBr$_3$ scintillator, although hygroscopic, 
are the best choice to absorb and measure hard X-ray photons as they have high Z 
and an excellent light yield, around 60 photons/keV compared to more standard crystals, 
like NaI, which produced only around 40 photons/keV or even less. 
However, LaBr$_3$, contrarily to CeBr$_3$, has a quite strong intrinsic background, 
which makes it hardly suitable for a high energy telescope.
 
In our previous study, a whole bulk of CeBr$_3$ scintillator crystal was used in the instrument model \citep{chang19}.
Although scattering location in the bulk can be determined to an accuracy of 3 mm or so \citep{gostojic16}, 
multiple scatterings in a single bulk cannot be separated and therefore will result in improper Compton event reconstruction. 
Using an array of bar scintillators, with each bar having a cross section size of 3 mm $\times$ 3 mm, can improve this issue.
We found that the detector efficiency can be increased by a factor of 5 or so if a bar array is used.

\begin{figure}[ht]
\plotone{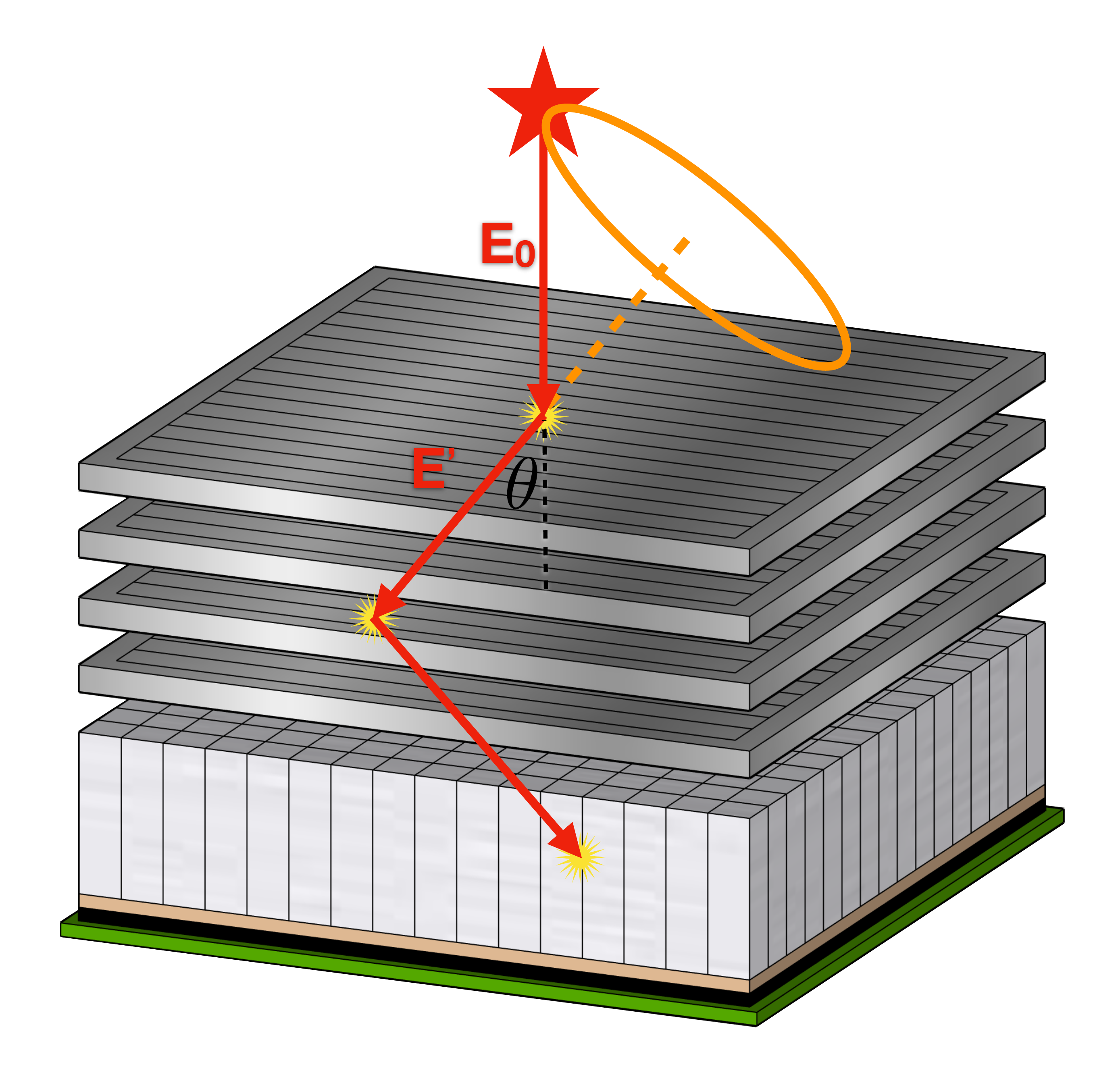}
\caption{An example of instrument models considered in our study. Shown here is Model 1-2. 
Four DSSD layers are on the top and a CeBr$_3$ scintillator array is at the bottom.
Beneath the scintillator array are a quartz window layer (yellow), the SiPM layer (black) and PCBs (green) that connect to SiPM.
A 1-mm-thick 5-side aluminum box, which together with the quartz window encloses the scintillator array, is not shown here for a better view of the scintillator array.
A 3-hit Compton sequence is shown for illustration: a photon of energy $E_0$ comes from the top, scatters with an electron
in the first silicon layer. A second scattering occurs in  the third layer and finally the photon is absorbed by CeBr$_3$ in the bottom. 
What are measured are the locations of the scattering/absorption, usually called a `hit', and the energy of recoiled/ionized electrons at each hit.
Temporal order of these hits is not known in the first place, but a good event reconstruction scheme may yield very successful determination
of the temporal order of the hits in one event (e.g. \citet{boggs00}) and the photon energy $E_0$ and the scattering angle $\theta$ can be known. The incoming direction of the photon can be determined only to the extent represented by the Compton cone. Once the first two hits in a Compton sequence are identified, the azimuthal scattering angle (not shown here) of a photon
coming from an assumed source direction can be determined. The distribution of this azimuthal scattering angle is then used for polarization analysis; see Section \ref{sec:mdp}.
\label{fig:m1-2}}
\end{figure}

\begin{figure*}[ht]
\plotone{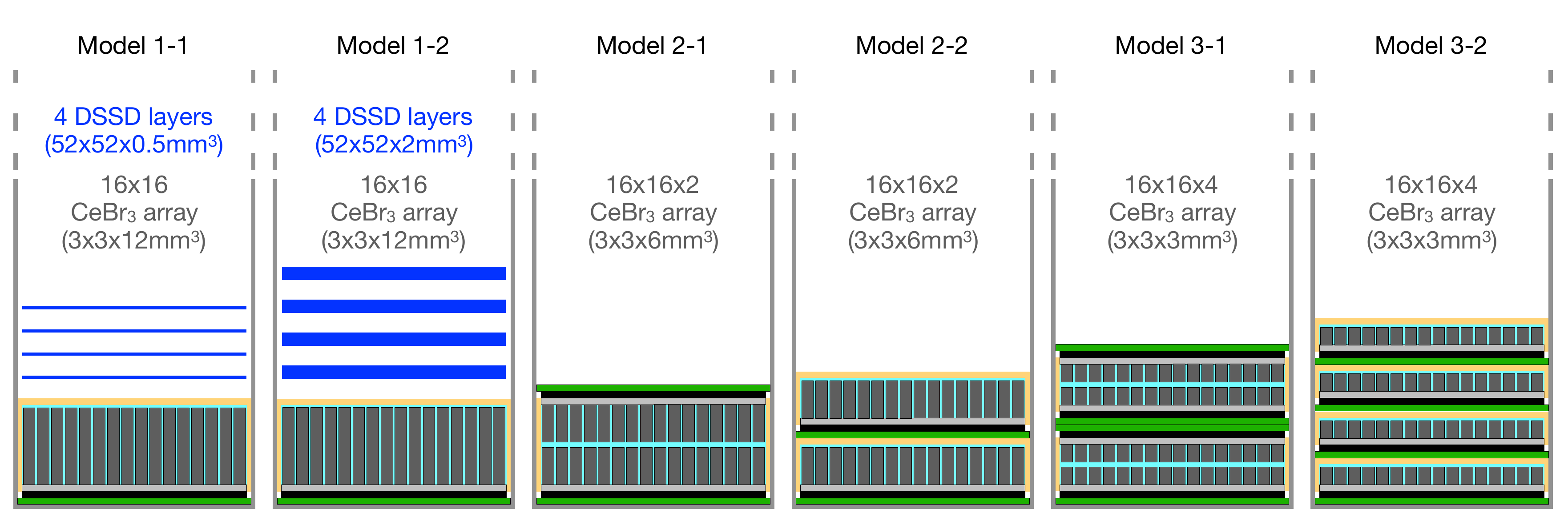}
\caption{Side view of the six instrument models studied in this work. 
The scale in the horizontal and vertical directions is not the same. 
To reveal details, different parts are not rigorously in scale either.
Sizes are indicated in the figure and are described in the main text. 
DSSDs of Models 1-1 and 1-2 are in blue. CeBr$_3$ scintillator bars are in dark gray.
Light blue lines stand for BaSO$_4$ wrapping for each scintillator bar and yellow lines for the 1-mm-thick aluminum box.
The gray thick lines surrounding each instrument are the 1-mm-thick 20-cm-long Pb shield.
The green/black/light-gray structures attached to each CeBr$_3$ scintillator array  
are PCB, SiPM and quartz windows, respectively. 
These are passive materials included in the simulation.  
\label{fig:allmodel}}
\end{figure*}

We consider six models, with two in a group.
Models 1-1 and 1-2 consist of four layers of double-sided silicon strip detectors (DSSDs) on the top as the scatter and 
a 12-mm-thick cerium bromide (CeBr$_3$) scintillator array at the bottom as the absorber. 
Each scintillator crystal is 3 mm $\times$ 3 mm $\times$ 12 mm in size and is wrapped with BaSO$_4$
on its top and four lateral sides to reflect the scintillation light. It is a 16 $\times$ 16 array, which can be readout from the bottom by four SiPMs, each with 8 $\times$ 8 readout channels
of corresponding size (3 mm $\times$ 3 mm) for one channel.  There is a 0.2 mm spacing between scintillator bars and between SiPM channels.
The width of each electrode strip on DSSD is 3 mm. There are 16 strips on one side, also with 0.2 mm space between adjacent strips.
The DSSD thickness is 0.5 mm for Model 1-1 and 2 mm for Model 1-2. The space between layers is 3 mm. 
These models are shown in Figure \ref{fig:m1-2} and Figure \ref{fig:allmodel}.
Since cerium bromide is hygroscopic,  scintillator arrays are, in our simulation, all coupled to a 1-mm-thick quartz optical window on the side to attach to SiPM, and are enclosed in a 1-mm-thick 5-side aluminum box.
In real manufacturing, the scintillator arrays will be hermetically sealed in such a box. 
These passive materials are included in our simulations.

The thickness, i.e., the length of each scintillator bar, is 12 mm in Models 1-1 and 1-2. 
It sets the spacial resolution in the detector volume for that direction to be worse than the other two directions.
It affects the Compton event reconstruction  accuracy. 
Besides, the total thickness of DSSD considered in Models 1-1 and 1-2
 seems small so that  many detected photons (we consider only Compton events of two or more hits)
 in fact only produce hits in the scintillator array 
 (92\% for Model 1-1, 72\% for Model 1-2 if 
 the incoming flux is from the top with a spectrum like that of Cyg X-1
 and the triggering energy threshold, which is discussed in the next section, is set at 40 keV).
These events have multiple hits in different scintillator bars in the same layer. 
There is no spatial resolution in the vertical direction. These hits are treated as though they all occur in the same plane.
It affects the performance of Compton event reconstruction. 
We therefore proceeded to try cerium bromide scintillator arrays with smaller thickness in our other models, 
in which only scintillator arrays are employed.
 These arrays are similar to that in Models 1-1 and 1-2, but with different thickness. 
This is similar to the instrument concept of COSI, which performs 3-D tracking of Compton scattering locations in the 
detector volume. The thickness of CeBr$_3$ can be reduced to improve
spacial resolution in the detector volume so that Compton event reconstruction can be better executed.
In Models 2-1 and 2-2,  two 6-mm-thick CeBr$_3$ arrays are used. The same SiPM described in Models 1-1 and 1-2 is used for readout. 
In Model 2-1, the top CeBr$_3$ array is read out
from the top so that the space between these two array layers can be minimized. In Model 2-2, the two layers are both read out from the bottom with a 4-mm space between the two layers
to allow the SiPM installation. Similar philosophy is applied to Model 3-1 and Model 3-2, both of which consist 
of 4 layers of 3-mm-thick CeBr$_3$ arrays. These models are all shown in Figure \ref{fig:allmodel}.

\section{Instrument performance} \label{sec:perform}

Throughout this study, we consider a higher triggering energy threshold at 40 keV and a lower one at 5 keV. 
The actual threshold value will be determined when the instrument
is built with all the readout electronics ready.
This threshold refers to a certain pulse height of the electric signal received by electrodes. There are 
mainly electric noises occurring all the time in the readout electronics with relatively low pulse height. To ignore those noises so that the electronics will not be occupied, a certain pulse height threshold is set for the electronics to further process only those pulses with pulse height higher than the threshold, which corresponds to a
certain triggering energy threshold.
The low threshold at 5 keV is probably feasible for DSSDs but not easy to achieve for CeBr$_3$ arrays,
whose threshold may need to be set at about 20 keV or above. 
In our study we set the same threshold for different sensors and use a lower one at 5 keV and a higher one at 40 keV
for a somewhat optimistic case and a more conservative one.
Any hit with energy of the recoiled/absorbed electrons lower than the threshold will not trigger the readout electronics.
Low energy hits in a Compton sequence are therefore  missed and result in improper event reconstruction or un-reconstructable events.
All these details are taken into account in our simulations.
We use Medium Energy Gamma-ray Astronomy Library (MEGAlib) \citep{zoglauer08} for all the simulations and analysis. 

\subsection{Detector Compton efficiency}\label{ssec:ef}
To study the on-axis detector Compton efficiency, we use input photon fluxes at different energies injected from the top in the simulation.  
The efficiency is defined as the ratio of the number of `useful' Compton events to the total number of photons passing through detectors. 
To find useful events among all the triggered ones, we exclude single-hit events, since they cannot be used for Compton reconstruction.
Events that have three or more hits in one single DSSD are also excluded, since the localization of hits through cross strips brings too much confusion for such cases. These two types of events can be recognized by readout electronics and therefore can be rejected on board
to reduce telemetry loading if desired. 
Events with pair-production hits are excluded for the reason of improper event reconstruction.  
We then further exclude events with eight or more hits and those which are un-reconstructable  because of incompatible kinematics in
the reconstruction attempt. 

\begin{figure*}[ht!]
\plotone{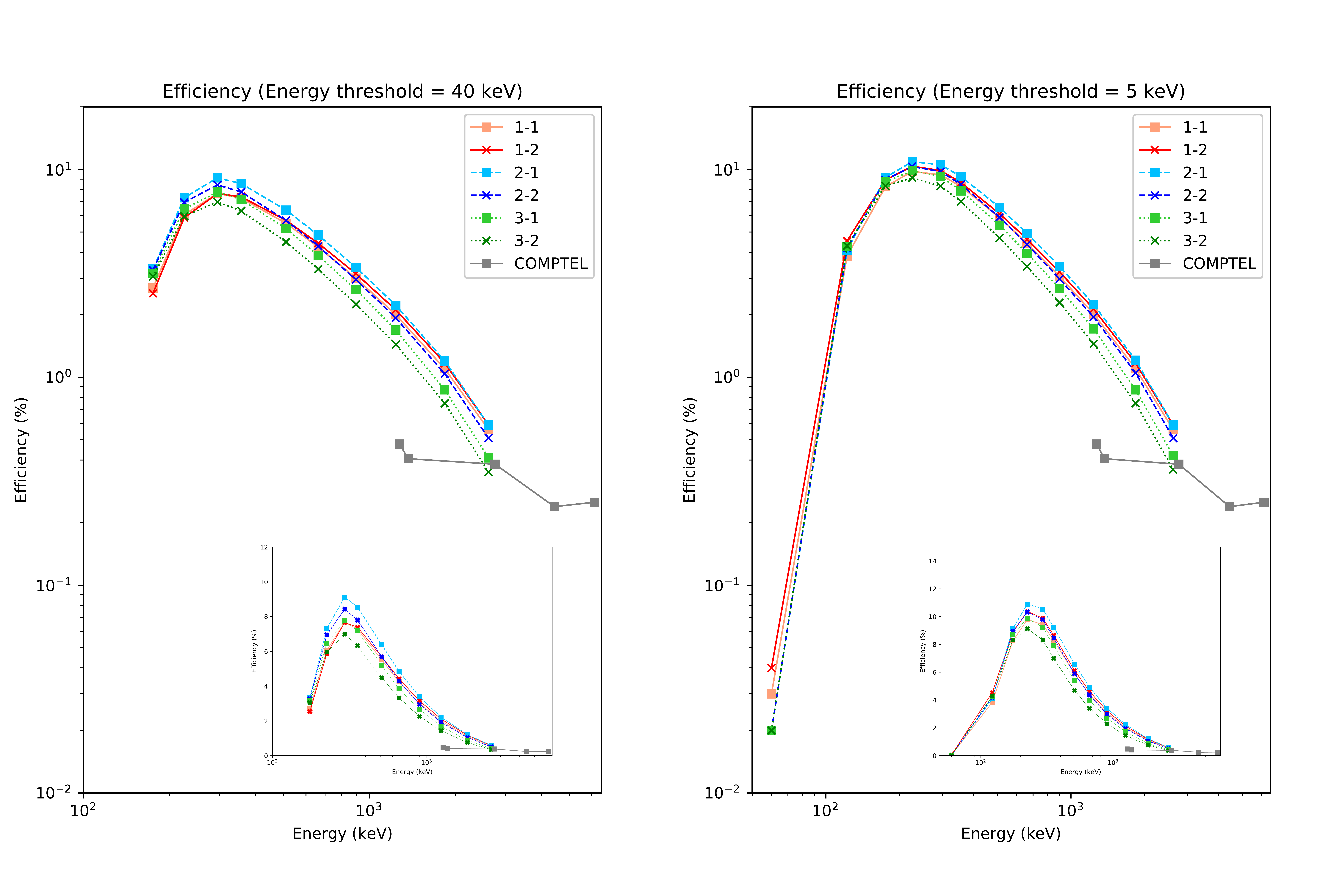}
\caption{Detector on-axis Compton efficiency of different models at different energies. 
The left panel is for the triggering energy threshold being set at 40 keV, and the right at 5 keV.
The two inserted figures show the same curves but with linear scale in the ordinate to make the difference among the six models more visible.
The COMPTEL efficiency is taken from \citet{schoenfelder93}.
 \label{fig:ef}}
\end{figure*}
For the efficiency discussed here, we apply an energy cut, that is, we consider only those events with reconstructed energy within the
3-$\sigma$ range from the photopeak, where $\sigma$ is the standard deviation determined from a Gaussian fit.   
We also apply an earth-horizon cut to exclude those events whose Compton cones are more than 50\% below the earth horizon. 
The on-axis efficiency of the six models is shown in Figure \ref{fig:ef}.
Although the difference is not really large, we can see that Model 2-1 has the highest efficiency at all the energies
except for the case of 5-keV threshold where Model 1-2 has higher efficiency at energies lower than 
about 160 keV.
This is because the scattering/absorption ratio is more favorable for silicon, given the atomic number, at this low energy.

Although with a better spatial resolution for hit localization, 
the Compton efficiency of Models 3-1 and 3-2 is the lowest at most of the energies among these models.
It is due to the presence of more passive materials between scintillator arrays.
Another cause is the relatively higher possibility for photons to escape out of the detector volume before completing a Compton sequence due to the larger vertical distance. 
These factors ail a proper Compton event reconstruction and result in a lower Compton efficiency. 
We note that photons of energy lower than about 160 keV will not produce more than one triggered hit 
if the triggering energy threshold is set at 40 keV. It is thus not possible to discuss efficiency at energies lower than that.
If the threshold is 5 keV, that energy is about 60 keV.

We also see that the efficiency of all the six models is higher than that of COMPTEL in the energy range from 1.5 MeV to 3 MeV.
In our instrument models, the acceptance angle, i.e., the range of scattering angles within which photons will not straightforwardly 
fly out of the detector volume,  is larger than COMPTEL's. 
There are multiple scattering events in our models, while COMPTEL essentially only picks up two-hit events. 
Roughly speaking, about half of  the Compton events detected in our models are with three hits or more.
Besides, the spatial resolution in the detector volume is much smaller than COMPTEL's.
All these factors help to yield more events with successful Compton event reconstruction
and therefore the Compton efficiency is enhanced. 
The instruments considered here, however, are small. 
Their capability to catch photons of higher energies is thus worse than COMPTEL.

\subsection{The shield}\label{ssec:sh}
In order to shield out most of the background, in particular the large number of low energy photons from the earth atmosphere,
we conducted a study to find a suitable shield.
We employed a low inclination (6 degrees), low altitude (575 km) earth-orbit radiation background model 
incorporated in MEGAlib for our simulation. 
It includes cosmic photons, protons, $\alpha$ particles, electrons and positrons, and albedo photons, protons, neutrons, electrons and positrons.
The dependence on incident directions (zenith angles) of each component is taken into account.
The energy range of the input background model is from 10 keV to 1 TeV.
We considered shields made of aluminum (Al), cesium iodide (CsI), and lead (Pb) with 1, 5, 10 mm thickness.
The CsI employed here was originally studied to evaluate a possible active shielding, but is anyway representative of a possible passive shielding of an intermediate atomic number.
The detector model used in this study was Model 1-2.
We compared the shielding effect on the reduction of the data to down link
and also the weight of the shield.
The results are shown in Table \ref{tab:shield}.
We can see that the reduction of data to down link (compared with the case of no shield) is greatest for
Pb shields, about 10\% -- 18\% for the case of 40 keV threshold and 43\% -- 50\% for the 5 keV case.
With weight consideration,  we decided to choose 1-mm Pb shield in our study.
\begin{table*}
\caption{Shielding study  \label{tab:shield}}
\begin{tabular}{lccccc} 
\hline\hline
Shield type & Triggered Events & Single-hit Events & Data to D/L & Reduction & Mass \\
 & (\%) & (\%) & (\%) & (\%) &  (kg)  \\
\hline
Bare & 100 (100) & 88.4 (95.3) & 8.83 (3.97) & - & - \\
Al,  1 mm  & 94.5 (51.3) & 83.0 (46.7) & 8.73 (3.87) & 1.07 (2.60) & 0.16 \\
Al,  5 mm  & 78.3 (30.5) & 67.1 (26.2) & 8.43 (3.56) & 4.55 (10.4) & 0.83 \\
Al,  10 mm & 65.6 (23.6) & 54.7 (19.5) & 8.58 (3.41) & 2.82 (14.2) & 1.76 \\
CsI, 1 mm  & 33.0 (12.4) & 21.5 (8.87) & 8.56 (2.74) & 3.07 (31.0) & 0.26 \\
CsI, 5 mm  & 22.7 (8.99) & 11.9 (5.90) & 7.94 (2.31) & 10.0 (42.0) & 1.38 \\
CsI, 10 mm & 20.8 (8.55) & 10.2 (5.57) & 8.17 (2.32) & 7.41 (41.5) & 2.93 \\
Pb,  1 mm  & 22.4 (8.73) & 11.5 (5.67) & 7.92 (2.27) & 10.3 (43.0) & 0.66 \\
Pb,  5 mm  & 18.3 (7.62) & 8.04 (4.80) & 7.27 (2.01) & 17.7 (49.5) & 3.47 \\
Pb,  10 mm & 17.5 (7.51) & 7.61 (4.82) & 7.41 (2.04) & 16.0 (48.6) & 7.39 \\
\hline
\end{tabular}
\tablecomments{We conducted simulations with a low-inclination, low altitude earth orbit background model taken from
MEGAlib to compare the down-link data volume with different shields. 
The instrument is pointing towards the zenith direction. The numbers in the
2nd, 3rd and 4th columns are the numbers of events normalized to the total trigger number of the
case without any shield (Bare) and are expressed in percentage. The instrument used in this
simulation is Model 1-2 and the simulation exposure time is 24 hours. It yields for the no shield case
about $1.1\times 10^7$ triggers with 40 keV threshold and $4.3\times 10^7$ triggers with 5 keV threshold. 
The events to down link are the triggered events after on-board rejection, which excludes single-hit events and 
events that have three or more hits in one single DSSD. 
What we concern most is the 4th column (Data to D/L) and the
5th column (Reduction), which shows the percentage of reduction in the data to down link compared
with the bare one. The 6th column shows the mass of the shield.}
\end{table*}

\begin{figure*}[ht!]
\plotone{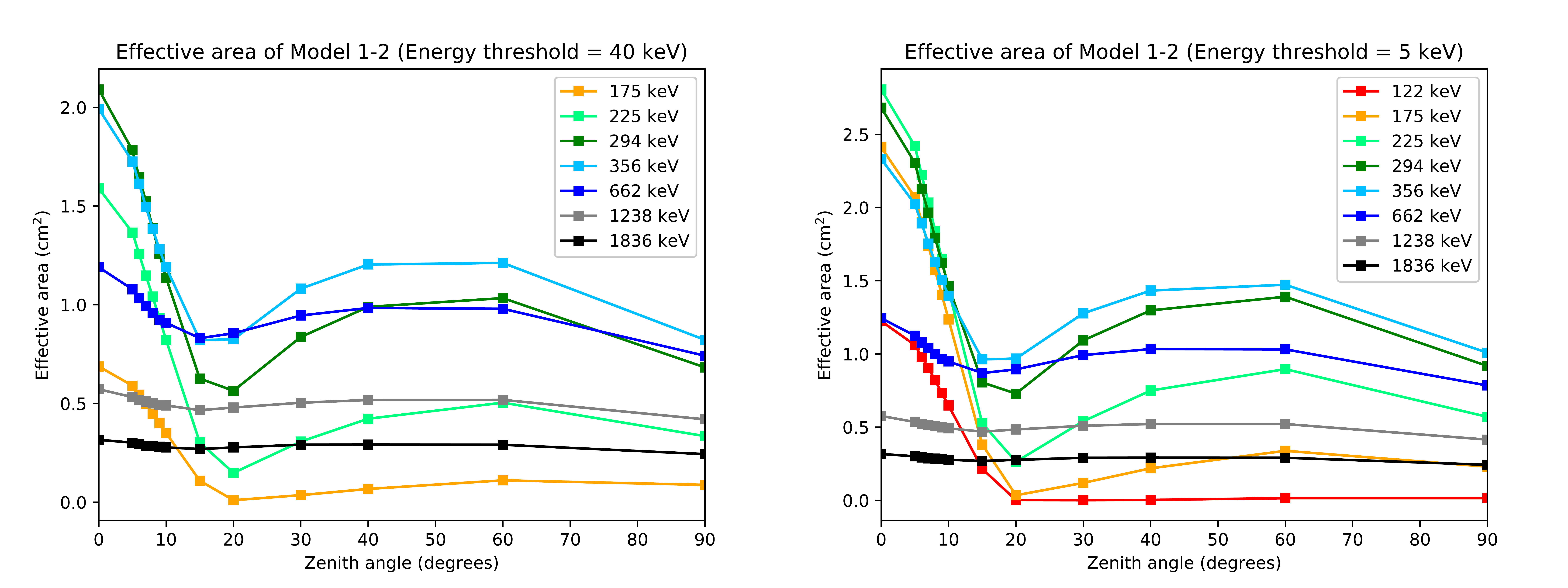}
\caption{Zenith-angle dependence of the effective area for different energies. Model 1-2 is employed in the simulation. The Pb shield is essentially transparent for high-energy photons.
The broad dip at zenith angle 15$^\circ$ -- 20$^\circ$ is due to the longer path in the shield for photons coming from that direction.
 \label{fig:sh}}
\end{figure*}

This shield in fact works only for photons of energy lower than about 300 keV. 
It is almost transparent for photons beyond about 600 keV, as shown in Figure \ref{fig:sh}, in which the zenith-angle-dependent  effective area is plotted for different energies.
The function of this shield is therefore only to reduce  low-energy background events.
From Figure \ref{fig:sh}, one can see that the FoV is about 30$^\circ$ in diameter for energy lower than 300 keV.
On the other hand, Cyg X-1 is persistent and so bright at energies higher than 100 keV that other sources 
(e.g. Cyg X-3 and GRS 1915+105 are about 10$^\circ$ and 25$^\circ$ away, respectively) 
may contaminate  the measurement only at the level of 10\% or so \citep{bouchet08, petry09}. 
V404 Cyg, which is about 5$^\circ$ away from Cyg X-1, may be much brighter than Cyg X-1 during its outburst, but that is only for a short
period of time.
Besides, the transparency  of this shield at high energy may also allow this instrument to act as a high-energy transient monitor. 
Although it has only a limited localization capability, which will be studied in more details in a future work, 
a constellation of such CubeSats may achieve good localization with the arrival-time-difference method. 

\subsection{Data rate in LEO and source detection}\label{ssec:rate}
For estimating telemetry demand of the payload, we derive the data rate from simulation with the 
aforementioned LEO radiation background.
To reduce the telemetry loading, on board rejection described in  Subsection \ref{ssec:ef} is adopted. 
The data rates (actually, event rates) for all the models are listed in Table \ref{tab:rate}. 
About half of the events, depending on energy, are two-hit events. Others have three hits or more.
It depends on future definition of data format to determine the data size. 
Roughly speaking, assuming 10 event/sec and 20 bite/event, science data from the instrument will be about 20 MB per day.
Of course this counts only the science data. The whole data volume will be significantly larger when housekeeping data is included.
\begin{table*}[t]
\caption{Data rate and signal to noise ratio \label{tab:rate}}
\begin{tabular}{lcccccc} 
\hline\hline
Model &  1-1 & 1-2 & 2-1 & 2-2 & 3-1 & 3-2 \\
\hline
\multicolumn{7}{c}{Triggering energy threshold 40 keV} \\
\hline
Data rate & 11.1 & 10.1 & 9.6 & 10.3 & 10.8 & 11.6 \\
Src rate & 3.80 $\times 10^{-2}$ & 3.86 $\times 10^{-2}$ & 3.52 $\times 10^{-2}$ & 3.42 $\times 10^{-2}$ & 3.15 $\times 10^{-2}$ & 3.03 $\times 10^{-2}$ \\
 & (2.72 $\times 10^{-2}$) & (2.80 $\times 10^{-2}$) & (2.57 $\times 10^{-2}$) & (2.54 $\times 10^{-2}$) & (2.42 $\times 10^{-2}$) & (2.37 $\times 10^{-2}$) \\
Bkg rate & 2.05 & 1.91 & 1.54 & 1.61 & 1.43 & 1.55 \\
 & (1.18) & (1.08) & (0.83) & (0.87) & (0.81) & (0.87) \\
1 Ms S/N & 26.5 & 27.9 & 28.4 & 27.0 & 26.4 & 24.4 \\
 & (25.0) & (26.9) & (28.1) & (27.3) & (26.8) & (25.4) \\
\hline
\multicolumn{7}{c}{Triggering energy threshold 5 keV} \\
\hline
Data rate & 11.9 & 11.3 & 10.4 & 11.1 & 11.7 & 12.6 \\
Src rate & 9.27 $\times 10^{-2}$ & 1.02 $\times 10^{-1}$ & 8.59 $\times 10^{-2}$ & 8.62 $\times 10^{-2}$ & 8.28 $\times 10^{-2}$ & 8.26 $\times 10^{-2}$ \\
 & (6.05 $\times 10^{-2}$) & (7.06 $\times 10^{-2}$) & (5.97 $\times 10^{-2}$) & (6.10 $\times 10^{-2}$) & (6.14 $\times 10^{-2}$) & (6.34 $\times 10^{-2}$) \\
Bkg rate & 2.52 & 2.44 & 1.97 & 2.07 & 1.86 & 2.01 \\
 & (1.28) & (1.23) & (0.99) & (1.04) & (1.00) & (1.09) \\
1 Ms S/N & 58.5 & 65.3 & 61.3 & 60.0 & 60.6 & 58.2 \\
 & (53.5) & (63.6) & (60.1) & (59.9) & (61.3) & (60.8) \\
\hline
\end{tabular}
\tablecomments{Data, source (Src) and background (Bkg) rates are all in units of event/s. See the main text for event selection criteria for these rates. 
Those numbers in parentheses are the results of further applying a 35-degree ARM cut.
The signal-to-noise ratio (S/N), which is simply the source counts divided by the square root of the background counts, with 1 Ms observation is very significant for all the models.
Listed in the upper part is the case of 40 keV triggering threshold and the energy range of selected photons is from 160 keV to 2 MeV.
In the lower part it is 5 keV threshold and photon energy from 60 keV to 2 MeV.}
\end{table*}

In Table \ref{tab:rate}, the source rate is derived from simulations with 
the Cyg X-1 spectrum  
obtained by INTEGRAL/IBIS observations during 2003 -- 2009 
\citep{laurent11} as the input from the zenith.
Cyg X-1 was mainly in the so-called low-hard state in that period \citep{rodriguez15}.
The source rate
includes only those events the same as in the consideration of detector efficiency in Section \ref{ssec:ef}, but
without energy cut around the photopeak,
since in reality we won't know the energy of the individual incoming photon. Those off-photopeak events give wrong 
information on energy and direction. To reduce them, usually, a spatial cut (an ARM cut; see Section \ref{sec:mdp} for more explanation) is applied, which keeps only those events with 'more correct' directions.  Even so, for spectral analysis, comparison of real data with simulations is required. 
Since the source is dim compared with the background, simulations of a much longer observation time
than the background case are required. 

To estimate the background rate, we exclude, from the down-link data,  those events with eight or more hits and those which are un-reconstructable  because of incompatible kinematics in
the reconstruction attempt. Events with pair-production hits are kept because that cannot be distinguished in a real measurement.
The incoming direction of a photon (represented by an event) is described by a Compton cone after a successful event reconstruction.
We further applied a 50\% earth horizon cut, which excludes events whose Compton cone has more than half
below the earth horizon, to both the background events and source events and then obtained 
the rates listed in Table \ref{tab:rate}. The earth horizon cut is crucial in reducing the background rate.
Throughout this study, the instrument is always pointing to the zenith direction. A significant fraction
 of the events due to the radiation background in orbit, in particular those from earth atmospheric albedo,  
can be removed with this earth horizon cut.  

We also examine the cases with the application of  a 35-deg ARM cut
(results shown in parentheses in Table \ref{tab:rate}). 
With this ARM cut, source and background events are both reduced, but the S/N value does not change much.
It indicates that the point spread function of this small instrument is quite extended and
Models 3-1 and 3-2 have a somewhat narrower point spread function than the other models
because of the smaller scintillator-bar size.
This can also be seen from the study of minimum detectable polarization degree
with different ARM cuts
in Section \ref{sec:mdp}.

We can see from Table \ref{tab:rate}, in fact, 
about 100-ks on-axis zenith-direction observation time  can  already yield a source detection at about 10-$\sigma$ level  for the 40 keV threshold case. 
For polarization measurement, of course, much longer observation time is needed.  
The detection discussed here is based on source counts divided by the square root of background counts. 
The shield (to reduce low-energy radiation in orbit)
and the earth horizon cut after Compton event reconstruction (to reduce events due to earth albedo),
significantly depress  the background rate. Our intention to check this detection level is to have a first check of whether it is feasible at all to measure polarization. Without a meaningful detection, polarization measurement is not possible. 

Nonetheless, the detection level shown here hints 
at a possibility to monitor Cyg X-1 flux on an integration time scale of days or a week. 
We note, however, the following caveats: 
(1) In our simulation, we consider only the case that the instrument is always pointing to the zenith and Cyg X-1 is always on-axis. It is of course not the case in real space flight. We have not yet simulated a real flight, 
which will definitely downgrade the detection level.
(2) The 1-mm Pb shield is not a collimator, but only to block low-energy photons, as shown in Figure \ref{fig:sh}. There are other relatively bright, nearby sources, such as Cyg X-3 and GRS 1915+105, which are about 10$^\circ$ and 25$^\circ$ away, respectively, and
may contaminate  the measurement at the level of 10\% or so \citep{bouchet08, petry09}. 
(3)  The S/N values in Table \ref{tab:rate} are based on photon counts in a broad energy range altogether. 
To distinguish detection in different energy band, or to perform spectral analysis, longer integration time 
to have more photons from the source is desired. 
 
\section{Minimum Detectable Polarization (MDP)}\label{sec:mdp}

\begin{figure*}[ht!]
\plotone{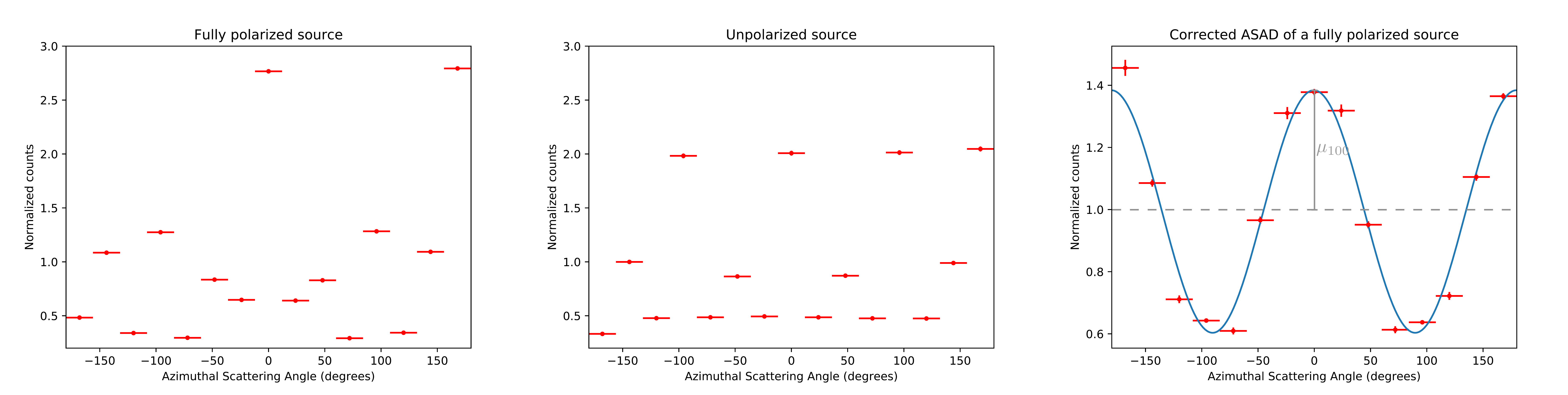}
\caption{One example of the azimuthal scattering angle distribution (ASAD) and the modulation factor.
These are from a 10-Ms simulation with conditions described in the main text. Shown in the left panel is the mean-normalized ASAD of an assumed 100\% polarized source with a polarization angle at 0 degree, which is defined to be the direction
perpendicular to one of the four surrounding sides of the instrument. In the middle is the ASAD of an assumed totally unpolarized source, 
and in the right the mean-normalized corrected ASAD of the 100\% polarized source, which is basically the curve in the left panel divided by the one in the middle. The $\mu_{100}$ shown in the right panel is the ratio
$A/B$ of that appearing in Eq.(\ref{eq:asad}).
For the unpolarized source (the middle panel), the azimuthal scattering angle preference bias at 0, $\pm$ 90, and 180 degrees, and, to a smaller extent, at $\pm$ 45 and $\pm$ 135 degrees, is due to the pixel-like structure of the sensor units.
Many successive hits happen in adjacent scintillator bars.  
 \label{fig:asad}}
\end{figure*}
Compton scattering is polarization dependent (e.g., \citet{lei97}). 
The probability density function of scattering into a particular azimuthal direction $\eta$ goes like (e.g., \citet{lowell17})
\begin{equation}\label{eq:polcos}
P( \eta) = A\cos\left(2( \eta-\eta_{0})\right) + B
\,\,\, ,
\label{eq:asad}
\end{equation}
where the amplitude $A$, offset $B$, and the polarization angle $\eta_0$ 
are parameters to fit the measured azimuthal scattering angle distribution (ASAD).
The modulation factor $\mu$, defined as $A/B$, can be compared with that of an assumed 100\% linearly polarized photon beam, 
usually denoted as $\mu_{100}$,
to estimate the polarization degree $\mu/\mu_{100}$ . 
Before using Eq.(\ref{eq:asad}) to fit an ASAD, one should subtract the background ASAD for cases of real measurements
and divide the background-subtracted ASAD
by a mean-scaled ASAD of an assumed totally unpolarized source to obtain a corrected ASAD,
which is then used for fitting with Eq.(\ref{eq:asad}) to find $\mu$. 

The ASAD of both totally polarized and unpolarized sources are obtained 
from simulations with proper settings.
In Figure \ref{fig:asad} we show an example of obtaining $\mu_{100}$ from a simulation.
In that simulation, the instrument model is Model 1-2, 
the incoming flux is from the top with a Cyg X-1 spectrum,
the exposure time is 10 Ms, 
the triggering energy threshold is 5 keV, 
a 50\% earth horizon cut is applied, and the photon energy considered is from 160 keV to 250 keV.
Besides the earth horizon cut, we further apply an ARM (angular resolution measure) cut at 35 degrees 
to the events to
produce the ASAD shown in Figure \ref{fig:asad}.

The ARM is the minimum angular distance from a Compton cone to the target direction 
(see, e.g., \citet{bandstra11}). A larger ARM indicates a lower chance of the photon (represented by the Compton cone in question) being really from the target. An ARM cut at 35 degrees means only those events
with ARM smaller than 35 degrees are kept for analysis.
In the following MDP discussion, we try ARM cuts at different angles.

The smallest polarization fraction which can be detected for a given source count rate $R_{S}$, 
background count rate $R_{B}$, 
and observation time $T$, 
is described by the minimum detectable polarization (MDP)\citep{weisskopf10}: 
\begin{equation}\label{eq:mdp}
\mathrm{MDP}=\frac{4.29}{\mu_{100}R_{S}}\sqrt{\frac{R_{S}+R_{B}}{T}} = \frac{4.29}{\mu_{100}}\frac{\sqrt{C_{S}+C_{B}}}{C_{S}}
\,\, ,
\end{equation}
where the factor 4.29 corresponds to a confidence level of 99\%, and $C_S$ and $C_B$ are the total counts from the source and the background, respectively.
We use Eq.(\ref{eq:mdp}) to derive MDP with $\mu_{100}$, $C_S$ and $C_B$ 
obtained from simulations for different instrument models in different energy ranges 
and with different ARM cuts.

\begin{figure*}[ht!]
\plotone{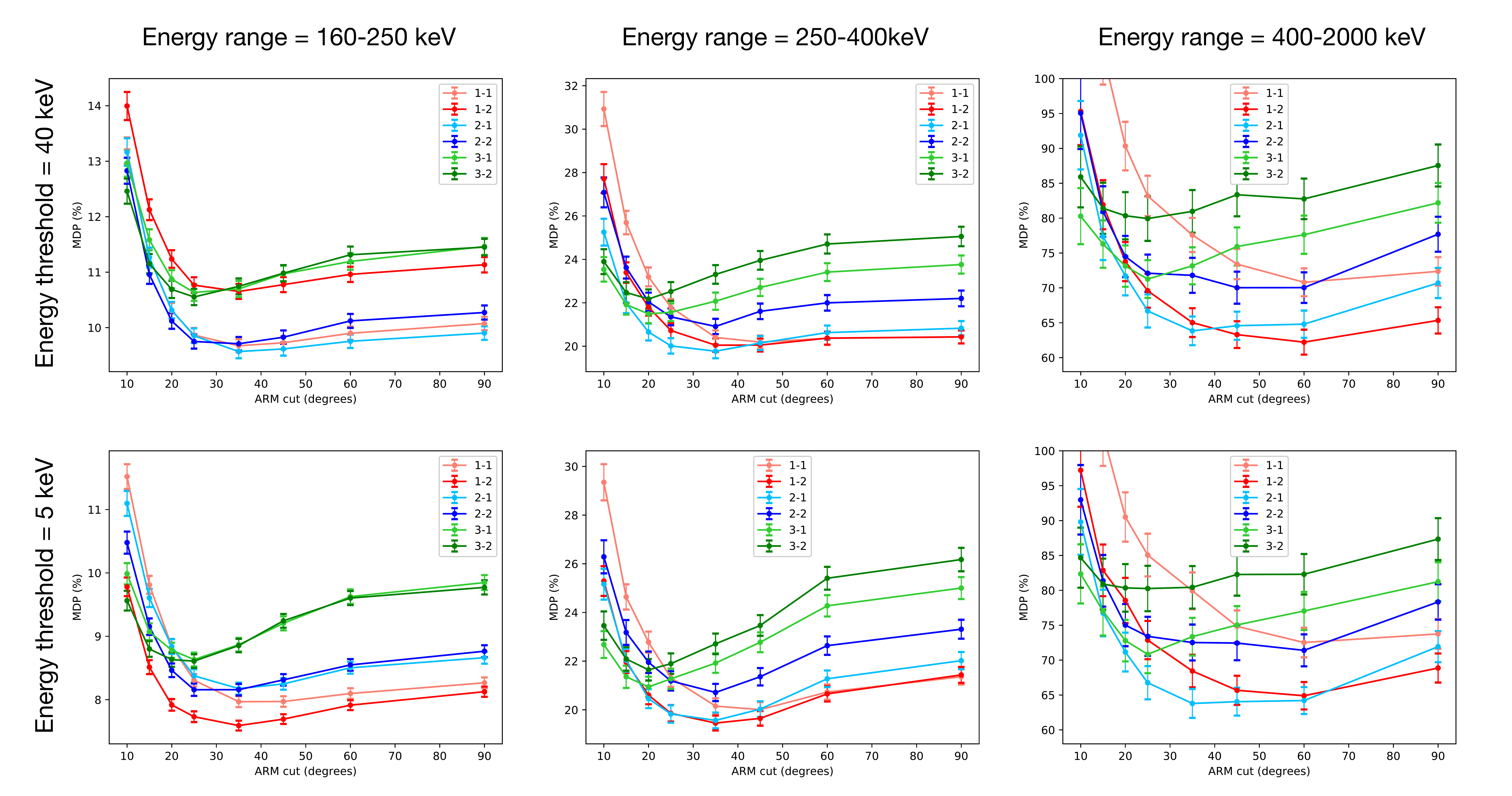}
\caption{The minimum detectable polarization (MDP) at different ARM cuts for the six models in three energy bands. 
The upper panels are for the case of 40-keV energy threshold and the lower ones are for 5-keV threshold.
\label{fig:mdp}}
\end{figure*}

In the discussion about the source rate and background rate in Section \ref{ssec:rate}, we apply only a 50\% earth horizon cut but not any
ARM cut for event selection. It is in fact not yet really optimized, although the detection significance is already quite high.
In exploring MDP, we check ARM cut at several different angles to find the optimal one.
Shown in Figure \ref{fig:mdp} is the MDP in the energy bands of 
160 -- 250 keV, 250 -- 400 keV, and 450 -- 2000 keV for a 10-Ms on-axis observation. 
The upper panels are with the 40-keV triggering energy threshold and the lower ones 5 keV. 
We can see in the upper panels that Model 2-1 can achieve the lowest MDP in these three energy bands, except that Model 1-2 is somewhat better (lower MDP) in 400 -- 2000 keV. 
The corresponding lowest MDP is about 9.5\% in 160 - 250 keV, 20\% in 250 - 400 keV, 
and 65\% in 400 - 2000 keV for Model 2-1. 

\begin{figure}[h]
\plotone{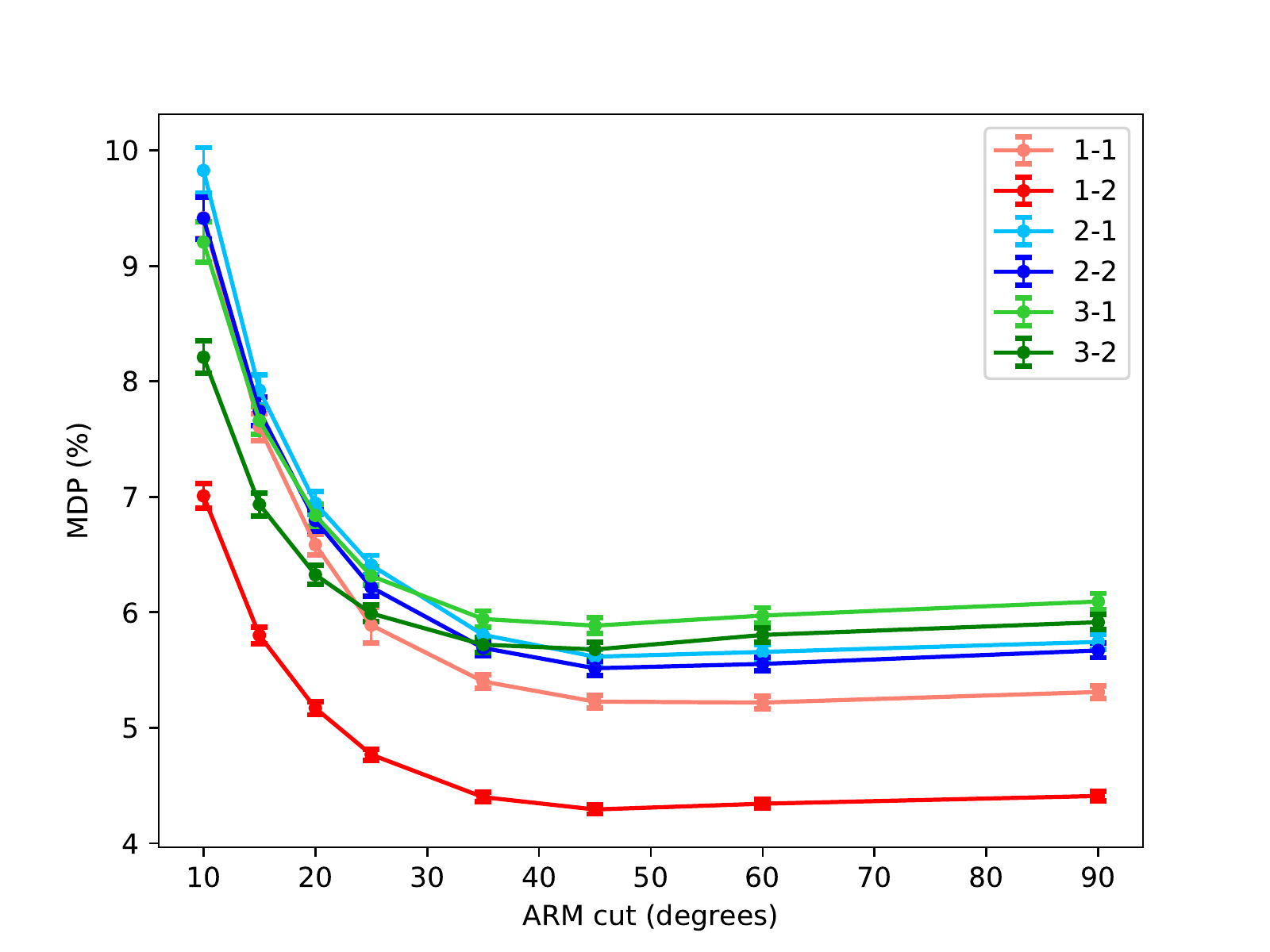}
\caption{The MDP in 60 -- 160 keV at different ARM cuts for the six models. The triggering energy threshold is set at 5 keV. Photons in this energy range do not produce any detectable Compton event in the instrument 
if the triggering energy threshold is 40 keV.
\label{fig:mdp-le}}
\end{figure}

In the lower panels, with a low threshold at 5 keV, Model 1-2 becomes the best, in particular in low-energy bands.
This is because 
silicon detectors are more sensitive to low-energy events and also inclined to produce low-energy hits in the Compton scattering  sequence. 
It is also related to the higher modulation of the differential Klein-Nishina cross section in the azimuthal scattering angle
for lower-energy photons.
The MDP is similar to that in the upper panels for 250 -- 400 keV and 400 -- 2000 keV, 
but goes down to about 7.5\% for 160 -- 250 keV.
This can be understood from the higher efficiency of Model 1-2 shown in Figure \ref{fig:ef} and the larger number of photons at low energy. 
As we mentioned earlier, this low triggering energy threshold is achievable probably only for DSSDs but not for CeBr$_3$.
We can only take this as a somewhat optimistic case.
With this low energy threshold, it is possible to discuss the MDP in 60 -- 160 keV,
in which Model 1-2 can achieve an MDP about 4.5\% (Figure \ref{fig:mdp-le}) for an ARM cut at 45$^\circ$. 
The difference in MDP at different ARM cuts is quite small if the ARM cut angle is larger than about 30 degrees.
It is because of the poor imaging capability of this small instrument. 

\section{Discussion}\label{sec:dis}

As we presented in the above, MDP lower than the currently measured polarization degree or upper limits thereof can be achieved with the model concept described in this paper, 
in particular with Model 1-2 and Model 2-1, for a 10-Ms 
on-axis zenith-direction observation time. 
If the CubeSat life time can be longer than one year, MDP can be even lower.
The 10\% MDP for the 160 -- 250 keV band, obtained with a conservative triggering energy threshold of 40 keV,
is lower than the current SPI upper limit, 20\% in 130 -- 230 keV \citep{jourdain12}. 
A lower triggering energy threshold, when achievable, will give even better results.
Polarization measurement in these energy bands will bring strong constraints to theoretical models for hard X-ray emissions from Cyg X-1. 
However, because of the large field of view (about 30 degrees in diameter for photon energy lower than about 300 keV; see Figure \ref{fig:sh}) and the transparency of the shield for photons of higher energies, 
contamination from other sources can be at the level of about 10\%
\citep{bouchet08,petry09}.

The 10-Ms observation time referred above is for an on-axis, zenith-pointing observation.
It is about 4 months. Considering some more realistic settings, such as the duty cycle in an orbit,  
Cyg X-1 zenith angle variation and more passive materials in a real spacecraft which have not yet been included in the simulation, a CubeSat life time of one or two years is probably needed.
With the fast advance of CubeSat technology and experiences in recent years, 
a longer life time should be achievable. Some examples include ASTERIA \citep{knapp18}, operating from Nov. 2017 to Dec. 2019, HaloSat \citep{kaaret19}, from Oct 2018 to June 2020 (planned), MinXSS-2 \citep{mason19}, launched in Dec 2018 for a 4-year mission, and PolarLight, launched in Oct 2018 and still in operation \citep{feng19,feng20}.

Among all the six instrument models, the performance difference in efficiency and in MDP is not really large.
Although Model 2-1 looks promising, it requires more readout channels than Model 1-2.
More precisely speaking, Models 1-1 and 1-2 have 384 readout channels (plus 8 from guard rings), Models 2-1 and 2-2 have 512 channels, and Models 3-1 and 3-2 have 1024.
Assuming less than 1 mW power consumption for each channel, these models demand a power of less than 1 W. 
This is manageable for a CubeSat. 
Model 1-2 may require too high a voltage for its 2-mm-thick DSSD. 
The MDP of Model 1-1, whose DSSD is 0.5-mm thick, 
can be used to  set the range of the change in MDP when a DSSD thinner than 2 mm is employed in the instrument. 
As for the weight consideration, the DSSD module of Model 1-2 
is less than 60 gram, the CeBr$_3$ module is less than 180 gram, and the 20-cm long Pb shield is about 660 gram.
Together with other components, such as the aluminum case enclosing CeBr$_3$, SiPM, PCB, etc, 
we expect the instrument to be
less than 1.5 kg. Other models are of a similar or smaller weight.
This weight is probably manageable if the instrument, which occupies roughly a 
$5 \times 5 \times 20$ cm$^3$ volume, is installed in a 2U space of a 3U CubeSat.

Besides pointing to Cyg X-1 whenever possible, the CubeSat should also rotate slowly along the axis of line of sight to eliminate possible systematic bias in azimuthal scattering angle measurement.
In order to obtain the background ASAD, 
blank-field observations are needed. 
These background observations may be conducted when Cyg X-1 is occulted by the earth. 

This instrument could  also be useful for measuring polarization of soft gamma-ray emissions from the Crab nebula and its pulsar. 
We will explore this possibility, as well as its performance of acting as a GRB monitor, either standing alone or in a constellation, in a future work.
 
\acknowledgments

We appreciate very much the anonymous referee's comments, which significantly improved this paper.
This work is supported by the Ministry of Science and Technology of the Republic of China (Taiwan) under the grant
MOST 108-2112-M-007-003.

%

\vspace{5mm}


\software{MEGAlib \citep{zoglauer08}
          }


\bibliographystyle{aasjournal}



\end{document}